# Competing Magnetic Phases in Li-Fe-Ge Kagome Systems

Zhen Zhang[1,†,*], Kirill D. Belashchenko[2,†], Xiaoyi Su[1,3,†], Atreyee Das[1,3], Sergey L. Bud'ko[1,3], Paul C. Canfield[1,3], Vladimir Antropov[1,3,*]

[1]Department of Physics and Astronomy, Iowa State University, Ames, IA 50011, USA

[2]Department of Physics and Astronomy and Nebraska Center for Materials and Nanoscience, University of Nebraska-Lincoln, Lincoln, NE 68588, USA

[3]Ames National Laboratory, U.S. Department of Energy, Ames, IA 50011, USA

[†]Equal contribution

[*]Email: zhenz1@iastate.edu (Zhen Zhang), antropov@iastate.edu (Vladimir Antropov)

## Abstract

Competing interlayer magnetic interactions in kagome magnets can lead to diverse magnetic phases, which enable various promising topological or quantum material properties. Here, the electronic structure and magnetic properties have been studied using first-principles calculations for the $LiFe_6Ge_6$, $LiFe_6Ge_4$, and $LiFe_6Ge_5$ compounds sharing the kagome $Fe_3Ge$ layer motif but with different interlayer arrangements. For $LiFe_6Ge_4$ and $LiFe_6Ge_5$, the predicted magnetic ground states are collinear antiferromagnetic (AFM) states involving a mix of ferromagnetic (FM) and AFM interlayer orientations. Whereas for $LiFe_6Ge_6$, an incommensurate cycloidal spin spiral is stabilized as a ground state, being close to a collinear A-type AFM state. The analysis of magnetic RKKY exchange coupling confirms the results of electronic structure calculations. The values of atomic magnetic moments are in good agreement with existing experimental estimations. Our experiments on $LiFe_6Ge_6$ single crystals have observed AFM ordering at ~540 K and transition to another magnetic phase, with a small FM component (possibly with spin canting), below ~270 K. Thus, our theory and experiment suggest the existence and sequence of collinear and noncollinear magnetic states in kagome $LiFe_6Ge_6$. Our findings provide a platform for exploring various novel magnetic phases and associated unconventional or topological magnetism.

## I. INTRODUCTION

The kagome lattice [1] and its three-dimensional realizations have attracted much interest in studying the interplay of band topology, electron correlations, magnetism, and various quantum phases [2]. The kagome lattice enables a variety of electronic structure features (e.g., Dirac cone, flat band, saddle point), and competitions between interplanar interactions can lead to diverse magnetic states [3–13]. Among numerous kagome materials [3], the $AT_6X_6$ compounds (where A is a group 1–4 metal, T is a $3d$ transition metal, and X is Ge or Sn) have attracted great attention because they exhibit a variety of magnetic properties [3–13] and electronic band structure peculiarities [5,14–16]. In contrast, related kagome materials with the same constituent elements but different compositions, $AT_6X_4$ and $AT_6X_5$ compounds [17–23], have not received much attention for their magnetic properties. Li-Fe-Ge is one of the known ternary systems for which $AT_6X_6$, $AT_6X_4$, and $AT_6X_5$ kagome materials exist. Despite having been synthesized [17,18,24] a long time ago, their magnetic properties were not measured.

Recently, magnetic measurements [25] have been conducted for $LiFe_6Ge_6$, $LiFe_6Ge_4$, and $LiFe_6Ge_5$, which were all found to be antiferromagnetic (AFM) at room temperature. Computationally, these systems have been studied with some conflicting results. The Materials Project [26] database includes only the ferromagnetic (FM) configuration. A density functional theory (DFT) study [27] reported the ground states for $LiFe_6Ge_6$ and $LiFe_6Ge_4$ to be FM and AFM, respectively. Another study [28] for $LiFe_6Ge_6$ reported its ground state to be A-type AFM. Whereas Mössbauer spectroscopy revealed the magnetic moments on Fe atoms of 1.1–1.3 $\mu_B$ in all three compounds [25], the Materials Project [26] indicates these moments to be 1.63–2.41 $\mu_B$, and Meschke et al. [27] reported even larger moments of 2.83–2.9 $\mu_B$.

Here, we theoretically determine the magnetic ground states for the three Li-Fe-Ge systems. Self-consistent total energy and magnetic exchange coupling calculations revealed the stability of different collinear AFM ground states in all these systems. Magnetic interactions are FM within the kagome layers, and interlayer interactions are either AFM or FM, depending on the distance separating the adjacent kagome layers. Further, we perform spin spiral calculations and find an incommensurate spin spiral state in $LiFe_6Ge_6$ that is slightly below the collinear AFM ground state. Our measurements on $LiFe_6Ge_6$ single crystals have observed AFM ordering at ~540 K and a change in magnetic state below ~270 K that could be a transition involving collinear and noncollinear magnetic phases.

## II. RESULTS AND DISCUSSION

### A. Crystal structures

The crystal structures of the three Li-Fe-Ge compounds, $LiFe_6Ge_6$, $LiFe_6Ge_5$, and $LiFe_6Ge_4$, are displayed in Fig. 1, where layers of atoms are indicated. The stereographic representation of the crystal structures is supplied in Fig. S1 of the Supplemental Material [29]. $LiFe_6Ge_6$ crystallizes in the space group $P6/mmm$, while $LiFe_6Ge_4$ and $LiFe_6Ge_5$ crystallize in the space group $R\bar{3}m$. In all three materials, Fe atoms form $Fe_3Ge$ kagome layers with an experimental in-plane Fe-Fe bond length of 2.53(1) Å [25]. The kagome layers are flat in $LiFe_6Ge_4$ and $LiFe_6Ge_5$ and slightly puckered in $LiFe_6Ge_6$.

In order to understand the formation of the structure of the three compounds, we need to start with hexagonal FeGe [30] (space group $P6/mmm$), which can be viewed as its parent compound. The FeGe structure is formed by the alternate stacking of $Fe_3Ge$ and $Ge_2$ layers. In $Fe_3Ge$ layers, Fe atoms occupy the vertices of a kagome lattice, and Ge atoms occupy the centers of the kagome hexagons. The Ge atoms in the $Ge_2$ layers are located above the centers of the small equilateral triangles in the kagome layers, forming a honeycomb lattice. There are no lateral shifts between the $Fe_3Ge$ layers. The Ge-base-centered hexagonal prisms formed by Fe atoms in the nearest kagome layers are unfilled in the FeGe structure, leaving a void at their centers. There is one such void per three FeGe formula units, and the voids form a primitive hexagonal lattice. The AFM FeGe has been reported to exhibit a charge density wave upon cooling [30].

In the $LiFe_6Ge_6$ structure [24,25], half of the voids are filled by Li atoms arranged in a rather unusual A/BC/A stacking pattern. Note that the Ge atoms in the $Fe_9Ge_3$ ($Fe_3Ge$ in a $\sqrt{3} \times \sqrt{3}$ cell) kagome layers above and below the Li atoms are significantly displaced away from the Li atoms. This large displacement is only possible if the voids above and below the Li atoms remain unfilled. This suggests that the occupation of two voids that are nearest neighbors along the vertical direction is energetically prohibitive. Ruling out such an occupation, the A/BC/A stacking pattern is one of the simplest possible orderings with the 50% void filling fraction.

There is a related $HfFe_6Ge_6$ structure [31] (space group $P6/mmm$), which is also based on the partial occupation of the voids of the parent FeGe compound by Hf atoms. In contrast with Li atoms in $LiFe_6Ge_6$, Hf atoms in $HfFe_6Ge_6$ are arranged in a fully layered ABC/-/ABC stacking pattern, in which all voids are filled in every second $Ge_2$ layer. Here again, no voids that are nearest neighbors along the vertical direction are filled at the same time.

Apart from the mentioned FeGe, $LiFe_6Ge_6$, and $HfFe_6Ge_6$ structures, there is also a related $Fe_3Ge$ crystal [32]. Unlike the FeGe, $LiFe_6Ge_6$, and $HfFe_6Ge_6$ structures that contain well-separated kagome layers, the $Fe_3Ge$ crystal consists of directly stacked $Fe_3Ge$ kagome layers only. Recently, the FM $Fe_3Ge$ crystal has been reported to exhibit a spin-reorientation transition upon cooling [32].

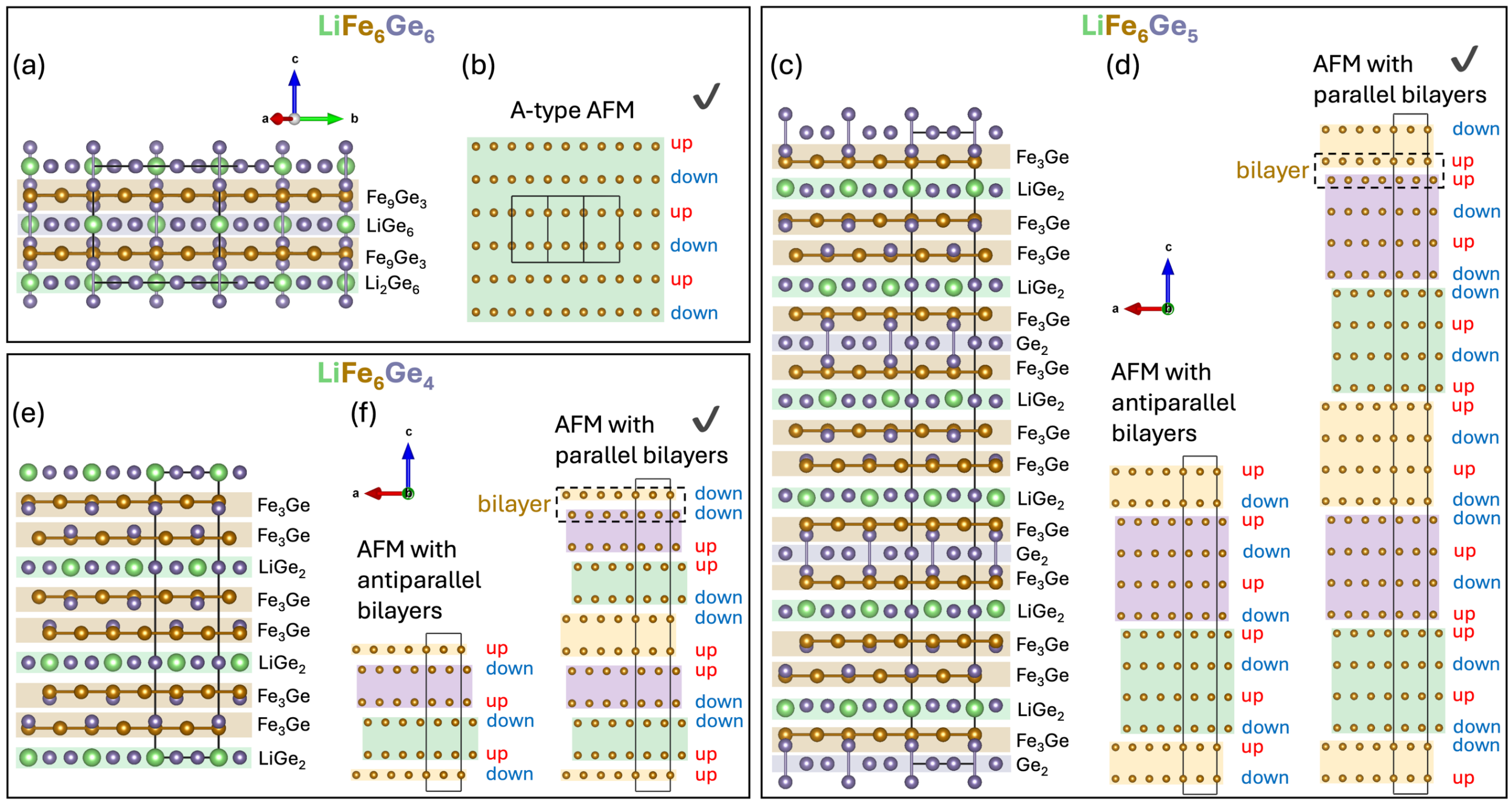

FIG. 1. Crystal structures of (a) $LiFe_6Ge_6$, (c) $LiFe_6Ge_5$, and (e) $LiFe_6Ge_4$, where layers of atoms are also indicated. Li, Fe, and Ge atoms are represented by green, brown, and purple spheres, respectively. Black rectangles indicate the 39-atom primitive cell of $LiFe_6Ge_6$, the 72-atom conventional cell of $LiFe_6Ge_5$, and the 33-atom conventional cell of $LiFe_6Ge_4$, respectively. Magnetic structures for AFM configurations of (b) $LiFe_6Ge_6$, (d) $LiFe_6Ge_5$, and (f) $LiFe_6Ge_4$. The collinear magnetic ground states are indicated by check marks.

The $LiFe_6Ge_4$ lattice [17,18,25] is based on the following stacking: $Fe_3Ge/Fe_3Ge/LiGe_2$, where the neighboring $Fe_3Ge$ kagome layers are displaced laterally with respect to each other, so that Ge atoms in each layer lie above or below some of the small equilateral triangles formed by Fe atoms in the neighboring kagome layer. The $LiGe_2$ layers are Ge honeycomb layers with Li atoms filling the center of each hexagon. Li atoms lie above and below the Ge atoms in the neighboring $Fe_3Ge$ kagome layers, while the placement of the Ge atoms in the $LiGe_2$ layers relative to the $Fe_3Ge$ layers is analogous to that in FeGe. The whole structure is arranged so that all atoms in the $LiGe_2$ layers form a primitive hexagonal lattice in which the Li atoms are stacked in the ABCA pattern. Each Ge atom in the $Fe_3Ge$ layers has exactly one Li atom directly above or below it in the adjacent $LiGe_2$ layer and is significantly displaced away from the Li atom.

The $LiFe_6Ge_5$ lattice [17,18,25] is even more complicated. The layer stacking sequence is $Fe_3Ge/Fe_3Ge/LiGe_2/Fe_3Ge/Ge_2/Fe_3Ge/LiGe_2$. The stacking of the adjacent $Fe_3Ge$ layers is the same as in $LiFe_6Ge_4$. The relative arrangement of the $Fe_3Ge$ and $Ge_2$ layers is analogous to FeGe, and that of the adjacent $Fe_3Ge$ and $LiGe_2$ layers is similar to $LiFe_6Ge_4$. All atoms in the $LiGe_2$ and $Ge_2$ layers form a (distorted) primitive hexagonal lattice in which the Li atoms are stacked as A0A-B0B-C0C-A0A, where the symbol 0 represents the unfilled void in a $Ge_2$ layer, and the dash represents the absence of a layer between a pair of adjacent kagome $Fe_3Ge$ layers.

Note that the $Fe_3Ge$ kagome layers separated by a $Ge_2$ layer (with or without some Li intercalation, i.e., $LiGe_6$, $Li_2Ge_6$, $LiGe_2$, or $Ge_2$) are always stacked directly on top of each other, without any lateral displacement, in all three compounds.

**B. Collinear magnetic ground states**

To determine the collinear magnetic ground states, we calculate the self-consistent total energies of the FM and various AFM configurations for the three phases, which are also shown in Fig. 1 next to their respective crystal structure. In $LiFe_6Ge_6$, the kagome layers are approximately, but not quite, equidistant. The monolayers stack above each other without a lateral shift, which is indicated by a uniform background color in Fig. 1(b).

As noted above, in $LiFe_6Ge_4$ and $LiFe_6Ge_5$, there are lateral shifts between the closely spaced kagome layers that are not separated by a $Ge_2$ layer; we will call a pair of such layers a bilayer. Figures 1(d) and 1(f) indicate a vertical (unshifted) block of kagome layers using a background color. Both materials have three kinds of blocks that are shifted laterally with respect to each other, shaded by three different colors. Bilayers appear at the boundaries between the blocks.

The AFM configurations for $LiFe_6Ge_4$ and $LiFe_6Ge_5$ are constructed so that the kagome layers inside the bilayers are either all parallel or all antiparallel, while the neighboring kagome layers inside the vertically stacked blocks are always antiparallel. To impose these magnetic orderings in periodic magnetic structures, $1 \times 1 \times 2$ supercells of both phases are required for the AFM configurations with parallel bilayers.

The total energies referenced from the most stable magnetic configuration are given in Table I, along with the magnetic moments on the Fe atoms. The collinear ground state of $LiFe_6Ge_6$ is A-type AFM, whose energy is 32.1 meV/Fe lower than that of FM, which is in good agreement with Wang [28] (27.1 meV/Fe). Flipping more and more neighboring layers to couple ferromagnetically or breaking the parallel alignment of spins within the kagome layers will increase the magnetic energy [28]. However, Meschke et al. [27] reported FM to be the ground state with the magnetization per Fe of 2.83 $\mu_B$. Our calculations yield the Fe magnetic moments of 1.68 and 1.69 $\mu_B$ in the A-type AFM, and 1.53 and 1.54 $\mu_B$ in the FM configuration. The two values correspond to two inequivalent Fe sites.

TABLE I. Total energy (meV per Fe) relative to the magnetic ground state (indicated in bold) and magnetic moment ($\mu_B$) on Fe atoms for various magnetic configurations. Two values for the magnetic moment correspond to two inequivalent Fe sites.

| $\Delta E$ | NM | FM | A-type AFM | AFM with antiparallel bilayers | AFM with parallel bilayers |
|---|---|---|---|---|---|
| $LiFe_6Ge_6$ | 269.3 | 32.1 | **0** | – | – |
| $LiFe_6Ge_5$ | 288.4 | 36.4 | – | 14.2 | **0** |
| $LiFe_6Ge_4$ | 338.3 | 33.2 | – | 22.6 | **0** |
| Moment | | | | | |
| $LiFe_6Ge_6$ | – | 1.53, 1.54 | **1.68, 1.69** | – | – |
| $LiFe_6Ge_5$ | – | 1.52, 1.96 | – | 1.72, 1.90 | **1.71, 1.97** |
| $LiFe_6Ge_4$ | – | 2.01 | – | 1.95 | **2.01** |

Both $LiFe_6Ge_4$ and $LiFe_6Ge_5$ have magnetic ground states in which the Fe atoms inside each bilayer are ferromagnetically aligned, while the kagome layers inside each vertical block are aligned antiparallel to each other. Thus, we conclude that the kagome layers separated by a $Ge_2$ layer have AFM interlayer coupling, while the two layers inside a bilayer are coupled ferromagnetically. The magnetic moments on Fe for the ground states of these phases are 2.01 and (1.71, 1.97) $\mu_B$ for $LiFe_6Ge_4$ and $LiFe_6Ge_5$, respectively. Note that Meschke et al. [27] reported a different AFM ground state for $LiFe_6Ge_4$.

Let us now compare our results with experimentally obtained magnetic moments. Mössbauer experiments [25] reported 1.1 $\mu_B$/Fe for $LiFe_6Ge_6$, 1.3$\mu_B$/Fe for $LiFe_6Ge_4$, and 1.2–1.3 $\mu_B$/Fe for $LiFe_6Ge_5$. Such values were obtained using the proportionality coefficient between $B_{hf}$ and the magnetic moment of 15 T/$\mu_B$. All these measurements were performed at room temperature.

If we instead use the coefficient for the Fe atom reported in Cadogan and Ryan [33] using the neutron-refined Fe moment, such an estimate would be different. For instance, with a coefficient of 10.4 T/$\mu_B$, the experimental moments would be 1.59 for $LiFe_6Ge_6$, 1.86 for $LiFe_6Ge_4$, and 1.73–1.86 for $LiFe_6Ge_5$, which are very close to our theoretical values of (1.68, 1.69), 2.01, and (1.71, 1.97), respectively. While the agreement is excellent, the experimental values should be larger at $T$ = 0 K, and the uncertainty in the hyperfine field factor could affect the comparison. Neutron diffraction measurements are desirable for establishing an unambiguous comparison.

For the found collinear ground states, magnetic anisotropy calculations indicate that in-plane magnetization is higher in energy than out-of-plane magnetization by 0.11 (0.10) meV/Fe for $LiFe_6Ge_6$, 0.24 (0.25) meV/Fe for $LiFe_6Ge_4$, and 0.16 (0.20) meV/Fe for $LiFe_6Ge_5$, respectively, obtained by GGA (and local density approximation (LDA)). Because these values are quite large and the magnetic moments are relatively small, the dipole-dipole contribution to magnetic anisotropy is expected to be negligible. Therefore, all three materials are easy-axis in DFT.

**C. Electronic structure**

The band structures of the collinear ground states are displayed in Figs. 2(a)–2(h). Both GGA and LDA bands are shown. Nonrelativistic and relativistic bands are displayed together for comparison. All of them are metallic. For $LiFe_6Ge_6$ (see Figs. 2(a) and 2(b)), the effects of spin-orbit coupling (SOC) become more pronounced along the out-of-plane Γ-A direction. Here, flat bands can be found at ~0.4 eV above the Fermi level $E_F$. A saddle point is located at the M point ~0.2 eV below $E_F$. In many magnetic kagome systems, flat bands are either not present (e.g., $Co_3Sn_2S_2$ [34], $Fe_3Sn_2$ [35]) or located far away from $E_F$ (e.g., FeSn [36]). A Dirac-point-like structure is found at the K point, which is slightly above $E_F$ in GGA and almost at $E_F$ in LDA. At this point, two bands closely interact with and repel each other, forming a local gap of 5.2 (8.4) meV by GGA (LDA) without SOC, while a third band crosses it (see Figs. 2(c) and 2(d)). Since there is an avoided crossing without SOC, it is not a real Dirac point.

For $LiFe_6Ge_4$ and $LiFe_6Ge_5$ (see Figs. 2(e)–2(h)), characteristics of the kagome electronic structure are not present near $E_F$. The distortion of the kagome electronic structure features [37] is likely caused by the interlayer hybridization within the bilayers, which includes all Fe atoms in $LiFe_6Ge_4$ but only some of them in $LiFe_6Ge_5$.

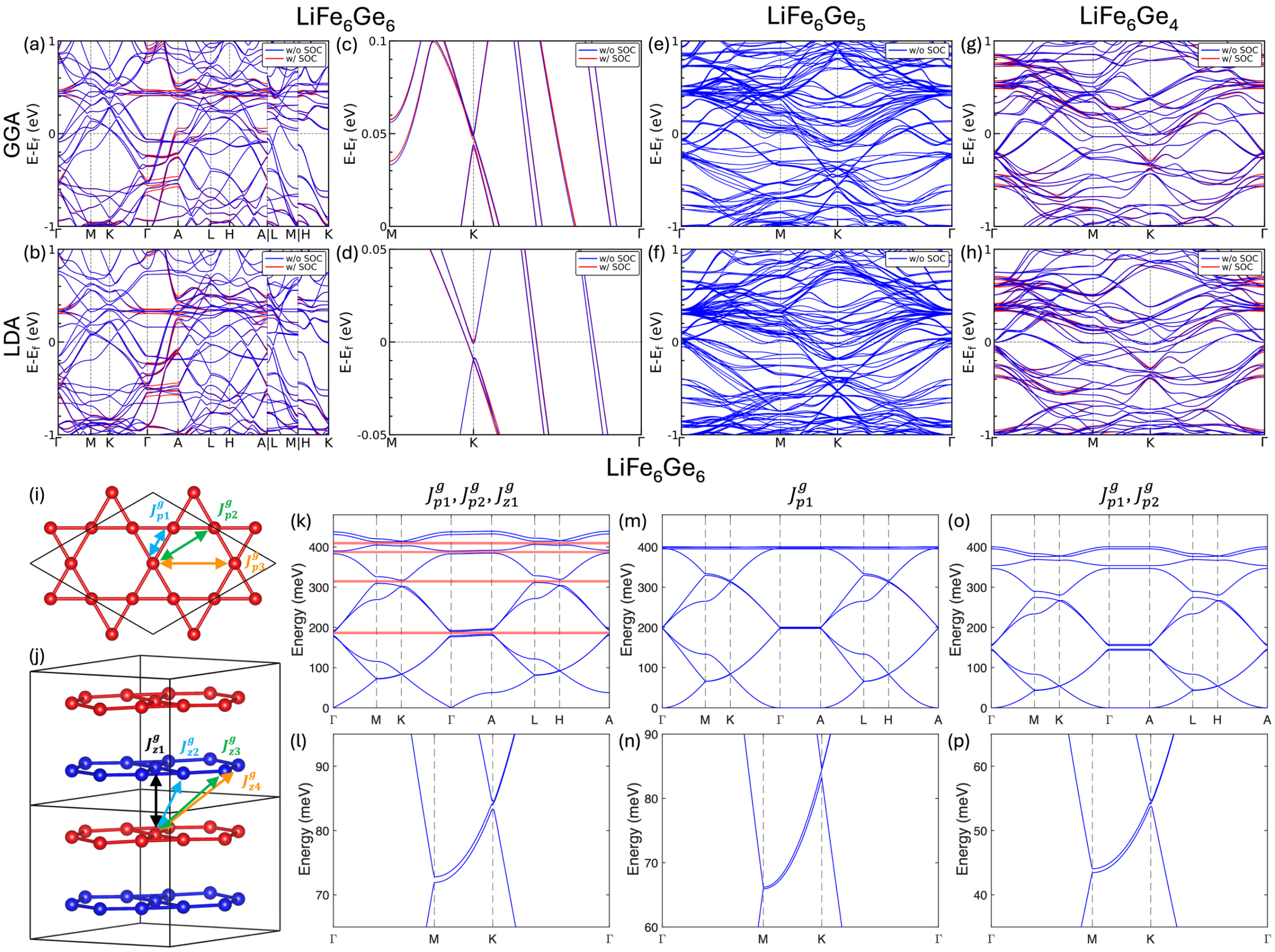

FIG. 2. Spin-polarized electronic bands of (a)(b)(c)(d) $LiFe_6Ge_6$, (e)(f) $LiFe_6Ge_5$, and (g)(h) $LiFe_6Ge_4$ obtained by (a)(c)(e)(g) GGA and (b)(d)(f)(h) LDA. Panels (c) and (d) are a zoom-in of panels (a) and (b), respectively, at the K point near the $E_F$. Blue and red curves indicate results obtained without and with SOC, respectively. k-points for $LiFe_6Ge_6$: Γ(0, 0, 0), M(1/2, 0, 0), K(1/3, 1/3, 0), A(0, 0, 1/2), L(1/2, 0, 1/2), H(1/3, 1/3, 1/2). k-points for the conventional cells of $LiFe_6Ge_5$ and $LiFe_6Ge_4$: Γ(0, 0, 0), M(1/2, 0, 0), K(1/3, 1/3, 0). Bands with SOC are not available for the large $LiFe_6Ge_5$ since accurate first-principles calculations for them are beyond our computational capability. (i) Intralayer and (j) interlayer magnetic interactions between the first several nearest neighbors in $LiFe_6Ge_6$. Blue and red indicate opposite spins. $J_{ij}$ are classified into groups, $J^g$. Within each group, $R_{ij}$ are approximately the same. The spin-wave spectra of $LiFe_6Ge_6$ obtained by (k) the leading $J^g_{p1}$, $J^g_{p2}$, and $J^g_{z1}$, (m) only $J^g_{p1}$, and (o) only $J^g_{p1}$ and $J^g_{p2}$. (l), (n), and (p) are zoom-ins of panels (k), (m), and (o), respectively, around the band structure along M-K below 100 meV. The red stripes in (k) indicate global band gaps.

**D. Magnetic exchange couplings**

To further confirm the observed structure-magnetism relationship and to rule out deviations from FM in-plane spin alignment, we calculated the magnetic exchange coupling parameters in the magnetic ground states. To better understand the magnetic interactions, inequivalent Fe sites and Fe-Fe distances for the in-plane and out-of-plane nearest neighbors (NN) in the three materials are displayed in Fig. S2 [29], where the calculated and experimental [25] Fe-Fe bond length values are also indicated in black and red, respectively. In $LiFe_6Ge_6$ (Fig. S2(a) [29]), due to the slight puckering of the kagome layer, there are two inequivalent Fe-Fe bonds with slightly different bond lengths, which are indicated in brown and blue. This leads to two inequivalent Fe sites, Fe1 (6i) with four blue bonds and Fe2 (12o) with two blue and two brown bonds. Both Fe1 and Fe2 have two slightly different out-of-plane Fe-Fe distances along the z-axis.

The Fe-Fe distances in $LiFe_6Ge_5$ and $LiFe_6Ge_4$ are indicated in a similar way. Although the kagome layers in these two phases are not puckered, they are distorted in the xy-plane. In $LiFe_6Ge_5$ (Fig. S2(b) [29]), there are two inequivalent Fe sites, Fe1 (18h) and Fe2 (18h). Fe1 atoms form the isolated kagome layer, while Fe2 atoms form a bilayer. In $LiFe_6Ge_4$ (Fig. S2(c) [29]), all Fe atoms (18h) are crystallographically equivalent. In both $LiFe_6Ge_5$ and $LiFe_6Ge_4$, the Fe-Fe distances in the xy-plane and along the z-axis inside the vertical blocks are analogous to those in $LiFe_6Ge_6$, while the intra-bilayer interlayer Fe-Fe bond lengths lie between the former two.

The calculated exchange parameters are listed in Tables S-I, S-II, and S-III of the Supplemental Material [29] for $LiFe_6Ge_6$, $LiFe_6Ge_5$, and $LiFe_6Ge_4$, respectively. These parameters are defined by the effective Heisenberg model, $E = -\sum_{ij} J_{ij}\, \hat{m}_i \hat{m}_j$ (see discussion near Equation 27 in Prange and Korenman [38]), where $\hat{m}_i$ are unit vectors, and each atomic pair is counted twice.

For $LiFe_6Ge_6$, to simplify the description, we grouped $J_{ij}$ with nearly identical $R_{ij}$ together as $J^g$. For ideal and equidistant kagome layers, the $R_{ij}$ of each group become identical. With this notation, the exchange parameters for several nearest in-plane ($J^g_{p1}, J^g_{p2}, J^g_{p3}$) and out-of-plane ($J^g_{z1}$, $J^g_{z2}, J^g_{z3}, J^g_{z4}$) neighbors are indicated in Figs. 2(i) and 2(j). According to Table S-I [29], only $J^g_{p1}$, $J^g_{p2}$, and $J^g_{z1}$ are significant. For Fe1, the intralayer parameters are $J^g_{p1}$ = 31.44 meV and $J^g_{p2}$ = -3.10 meV; for Fe2, they are $J^g_{p1}$ = (31.44, 31.14) meV and $J^g_{p2}$ = (-4.64, -3.65) meV. The leading positive $J^g_{p1}$ dominates the intralayer FM interaction, so that no in-plane magnetic frustration or noncollinearity is expected. The interlayer parameters are $J^g_{z1}$ = (-7.16, -11.33) meV for Fe1 and $J_{z1}^{g}$ = (-7.55, -10.19) meV for Fe2. The negative signs of $J_{z1}^{g}$ agree with the interlayer AFM configuration established above by the total energy calculations.

To determine the total coupling $J_{tot}^{iq}$ of a given atom *i* to an entire kagome layer $q$, we add up all the relevant exchange parameters: $J_{tot}^{iq} = \sum_{j \in q} J_{ij}$. For the Fe1 atoms, the intralayer and interlayer $J_{tot}$ are 107.70 and -27.73 meV, respectively; for Fe2, they are 104.76 and -29.59 meV. This further confirms the A-type AFM ground state for $LiFe_6Ge_6$. The total magnetic stability parameter between a given atom and the rest of the crystal is given by $J_0 = \sum_j J_{ij}\, \widehat{m}_i \widehat{m}_j$. The Néel temperature $T_N$ is estimated at 1040 K using $T_N = (2/3) J_0 / k_B$ within the mean-field approximation (MFA). A corresponding Monte-Carlo estimate should be expected to be 20–30% lower.

Figure 2(k) shows the spin-wave spectrum for $LiFe_6Ge_6$, which was calculated using the leading $J_{p1}^{g}$, $J_{p2}^{g}$, and $J_{z1}^{g}$ terms. Four global band gaps are present in the spectrum, which are indicated in red. The lower edge of these gaps is located at energies of 183.07, 311.66, 384.64, and 406.28 meV, with a gap size of 6.65, 5.97, 5.32, and 6.39 meV, respectively. A zoom-in of the spectrum below 100 meV along M-K is exhibited in Fig. 2(l). At the K point, analogous to the electronic structure shown in Figs. 2(a)–2(d), two cone-like bands closely interact with each other, forming a local gap of 1.33 meV, and a third band crosses the interacting point. This third band also interacts with the lower band along M-K with a split of 0.88–0.58 meV.

To trace back to the origin of these gaps, spin-wave spectra calculated using only $J_{p1}^{g}$ (Fig. 2(m)) and only $J_{p1}^{g}$ and $J_{p2}^{g}$ (Fig. 2(o)). Their zoom-ins below 100 meV along M-K are shown in Figs. 2(n) and 2(p), respectively. Note that $J_{p1}^{g} > 0$, $J_{p2}^{g}$ and $J_{z1}^{g} < 0$. By comparison, it is seen that the four global gaps primarily result from the addition of $J_{p2}^{g}$ (Fig. 2(o)), i.e., the (grouped) in-plane second nearest-neighbor couplings, while other couplings do not significantly contribute to these global gaps. In terms of the local gap/splitting along M-K below 100 meV, although $J_{p1}^{g}$, $J_{p2}^{g}$, and $J_{z1}^{g}$ all contribute to the gap/splitting size, the subtle feature of this band structure is already determined at the level of $J_{p1}^{g}$ (Fig. 2(n)), i.e., the (grouped) in-plane nearest-neighbor couplings.

The $J_{ij}$ calculations also confirm the magnetic ground states for the other two compounds. In $LiFe_6Ge_4$ Table S-III [29], the leading terms are $J_{ij}$ = (47.39, 36.63) meV and the summation $J_{tot}$ = 106.54 meV for the intralayer FM interactions; and $J_{ij}$ = -17.00 meV and $J_{tot}$ = -16.38 meV for the interlayer AFM interactions within the vertical blocks. In addition, for the intra-bilayer interlayer interactions, $J_{tot}$ = 13.11 meV confirms the preferred FM alignment within the bilayers. Therefore, the ground state of $LiFe_6Ge_4$ is confirmed to be the AFM configuration with parallel bilayers (see Fig. 1(f)). Its $T_N$ is estimated at 1050 K in MFA.

In $LiFe_6Ge_5$ Table S-II [29], analogous to $LiFe_6Ge_4$, the leading terms of $J_{ij}$ and the $J_{tot}$ confirm the intralayer FM interactions and the interlayer AFM interactions within the vertical blocks. In addition, $J_{tot}$ justifies the anticipated intra-bilayer interlayer FM interactions. Thus, the ground state of $LiFe_6Ge_5$ is confirmed to be AFM with parallel bilayers (see Fig. 1(d)). Its $T_N$ is estimated to be 990 K by the MFA.

Our DFT calculations of collinear magnetism reveal a clear and common relationship between magnetic ordering and Fe-Fe distance in this family of Li-Fe-Ge kagome systems: The magnetic ordering is FM for kagome layers in direct contact with each other (without a separating $Ge_2$ layer) and AFM for kagome layers separated by a $Ge_2$ layer, regardless of the amount of intercalation by Li.

**E. Spin spiral states**

We now examine the possibility of long-period or incommensurate magnetic structures in Li-Fe-Ge compounds. The analysis of the pair exchange coupling parameters described above shows that the Fe atoms within a given $Fe_3Ge$ kagome layer are strongly coupled ferromagnetically. Therefore, we will assume that each such layer is uniformly ordered but allows for coplanar spin spirals with arbitrary angles between the spins in adjacent layers. Using the generalized Bloch theorem [39], such spin spirals with arbitrary interlayer angles can be studied using first-principles calculations in the nonrelativistic limit.

As explained above under *Section IIA*, the $LiFe_6Ge_6$ structure may be obtained by inserting Li atoms into some of the voids within the honeycomb $Ge_2$ layers, in the parent hexagonal FeGe compound, in the A/BC/A stacking pattern. Therefore, in contrast to FeGe, where exchange interactions between each pair of nearby $Fe_3Ge$ layers are identical by symmetry, in $LiFe_6Ge_6$, the interactions between the $Fe_3Ge$ layers separated by $Li_2Ge_6$ and $LiGe_6$ layers are different. Thus, we consider the general case of a planar spin spiral in which the spins in the $Fe_3Ge$ layers separated by a $Li_2Ge_6$ layer make an angle $\pi + \alpha$ between each other, and those separated by $LiGe_6$ make an angle $\pi + \beta$. (The shift by $\pi$ is used because the lowest-energy collinear state has an AFM interlayer spin alignment.) The spiral angle across the crystallographic unit cell is thus $\alpha + \beta$. A demonstration of the magnetic structure in the spin spiral calculations is displayed in Fig. 3(a). Spin spiral calculations were performed to DFT self-consistency. There are two inequivalent Fe atoms in the $LiFe_6Ge_6$ structure; the local moments on these atoms were almost identical, within a narrow range of 1.53 to 1.55 $\mu_B$, for all values of $\alpha$ and $\beta$.

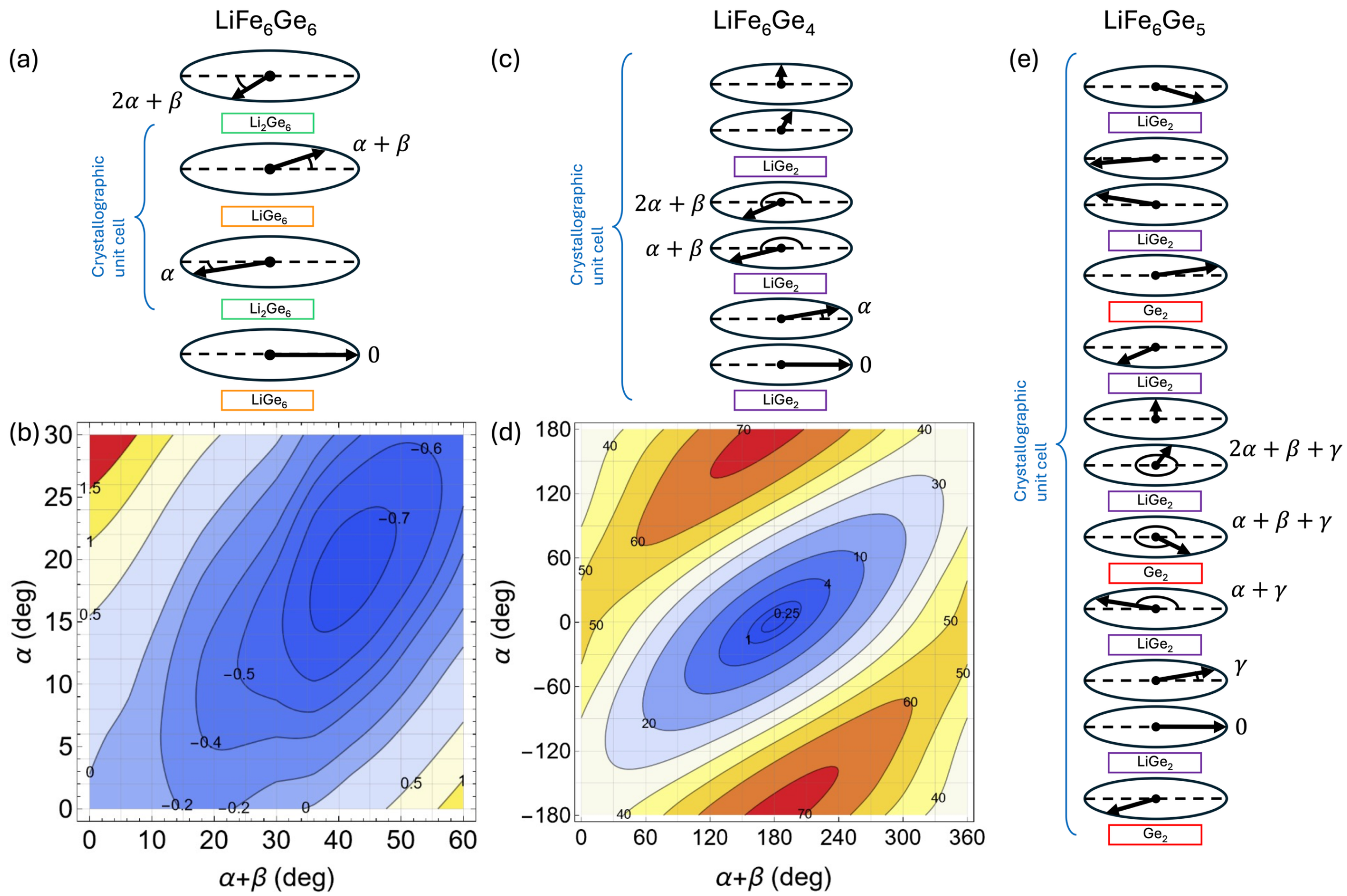

FIG. 3. Demonstration of the magnetic structures in the spin spiral calculations for (a) $LiFe_6Ge_6$, (c) $LiFe_6Ge_4$, and (e) $LiFe_6Ge_5$. Circles represent kagome layers. Arrows indicate the direction of FM spins within a kagome layer. Note that this demonstration does not imply that spin textures are helical spirals. In $LiFe_6Ge_6$, kagome layers are separated by a $Li_2Ge_6$ or $LiGe_6$ layer. In $LiFe_6Ge_4$, kagome layers are separated by a $LiGe_2$ layer. In $LiFe_6Ge_5$, kagome layers are separated by a $LiGe_2$ or $Ge_2$ layer. Total energy (in meV per Fe site) in (b) $LiFe_6Ge_6$ and (d) $LiFe_6Ge_4$, relative to the lowest-energy collinear spin configuration, as a function of the spiral angles. See text for more details on the definition of the angles $\alpha$, $\beta$, and $\gamma$.

Figure 3(b) shows the dependence of the calculated total energy on the angles $\alpha + \beta$ and $\alpha$ relative to the AFM spin configuration. The minimum energy ($-0.74$ meV/Fe) is reached in the spiral state with $\alpha + \beta \approx 41.6°$ and $\alpha \approx 19.2°$. Thus, the angles $\alpha$ and $\beta$ are very similar, suggesting the interlayer couplings across the $Li_2Ge_6$ and $LiGe_6$ layers are similar. The relatively small stabilization energy for the spin spiral state suggests that the phase transition to the spiral structure happens at a rather low temperature. However, the magnetic entropy could significantly increase the possible temperature of this transition between two magnetically ordered phases.

According to *Section IIB*, the magnetocrystalline anisotropy (MCA) in the lowest-energy collinear $LiFe_6Ge_6$ phase ($\alpha = \beta = 0$ in our notation) is easy-axis. With such an MCA, the coplanar spin spiral is expected to adopt the cycloidal configuration in which the magnetic moments lie in a plane containing the (easy) hexagonal axis. Indeed, it is reasonable to assume that the MCA is dominated by "single-layer" terms, in the sense that $E_{MCA} = -KN_l \sum_l m_{lz}^2$, where $K > 0$ is the easy-axis anisotropy constant (per Fe atom), $\mathbf{m}_l$ is the unit vector designating the orientation of the magnetic moments in the $l$-th kagome layer, and $N_l$ is the number of Fe atoms per layer. This assumption allows the exchange anisotropy terms to contribute to the MCA as long as the two sites belong to the same kagome layer. Consider now an incommensurate coplanar spin spiral defined by a unit vector $\mathbf{n}$ orthogonal to all the spins. In such a spiral, the unit vectors $\mathbf{m}_l$ are uniformly distributed over the unit circle orthogonal to $\mathbf{n}$. Averaging over that unit circle, we find $E_{MCA}/N = -\frac{1}{2} K \sin^2 \theta$, where $\cos\theta = n_z$ and $N$ is the number of Fe atoms in the crystal. Evidently, $E_{MCA}$ is minimized at $\theta = \pi/2$, which corresponds to a cycloidal spin spiral. Naturally, in the easy-plane case $K < 0$, the anisotropy energy $E_{MCA}$ is minimized by the helical spiral in which all spins lie in the easy plane.

In the $LiFe_6Ge_4$ compound, as noted above under *Section IIA*, $Fe_3Ge$ kagome bilayers are separated by $LiGe_2$ layers. Thus, a generic planar spin spiral, just as in $LiFe_6Ge_6$, may be described by two angles. Here, we denote the angle between the spins in a kagome bilayer as $\alpha$, and the angle between the vertically stacked kagome layers across a $LiGe_2$ layer as $\beta$ (see Fig. 3(c)). The magnetic moments on all Fe atoms vary between 1.84 and 1.98 $\mu_B$ depending on these angles. The calculated total energy profile for $LiFe_6Ge_4$ is shown in Fig. 3(d). Here, the minimum energy corresponds to a collinear state with $\alpha = 0$ and $\beta = \pi$, i.e., the spins in the adjacent kagome layers are parallel to each other, and those across a $LiGe_2$ layer are antiparallel. This configuration agrees with that determined above under *Section IIB* using collinear total-energy calculations. The shape of the energy profile in Fig. 3(d) shows that the FM exchange coupling between the adjacent kagome layers is somewhat weaker compared to the AFM coupling across a $LiGe_2$ layer. This observation is consistent with the magnetic exchange coupling calculations.

As noted above under *Section IIA*, the structural layers in $LiFe_6Ge_5$ form the KKAK0KAKKBK0KBKK… stacking sequence, where K denotes kagome $Fe_3Ge$ layers, A, B, and C the $LiGe_2$ layers with different sets of voids occupied by Li, and 0 the $Ge_2$ layers. A generic planar spiral in this structure is characterized by three different angles: $\alpha$ between the K layers in the KAK, KBK, or KCK blocks (which are the same, by symmetry), $\beta$ in the K0K blocks, and $\gamma$ in the adjacent layers within the KK bilayers (see Fig. 3(e)). The lowest-energy collinear structure established in *Section IIB* has $\alpha = \beta = \pi$ and $\gamma = 0$. We have performed self-consistent spin-spiral calculations allowing for $\alpha$, $\beta$, and $\gamma$ to deviate from these values independently by up to 50° in either direction, in steps of 10°. The collinear structure with deviations $\delta_\alpha = \delta_\beta = \gamma = 0$ has the lowest energy, and the energies of all calculated spirals with $\left(\delta_\alpha^2 + \delta_\beta^2 + \gamma^2\right)^{1/2} \leq 50°$ are well fitted to a positive-definite quadratic form. The lowest eigenvalue of the quadratic form is 0.00065 meV/deg$^2$, corresponding to the eigenvector (-0.086, 0.996, 0.03). This eigenvector corresponds to the varying $\beta$ angle with a small admixture of $\delta_\alpha$ and $\gamma$.

**F. Susceptibility, resistance, and Mössbauer measurements for $LiFe_6Ge_6$**

To address some band-structural findings, single crystals of $LiFe_6Ge_6$ were synthesized, as exhibited in Fig. 4(a), and their magnetic properties were measured. Figure 4(b) presents anisotropic temperature-dependent dc susceptibility measurements between 1.8 and 650 K. Ac susceptibility measurements in zero dc bias field are shown in Fig. 4(c). M(H) curves measured at selected temperatures are plotted in Figs. 4(d) and 4(e). These datasets taken together suggest the following evolution of magnetic properties. The feature at 540 K is a signature of an AFM transition. The anisotropic susceptibility behavior between ~275 K and ~540 K suggests that the moments in this temperature range are confined to the ab-plane. Below ~270 K, a second, broad, continuous transition to another magnetic phase is observed. The maximum in the real part of ac susceptibility is located at ~230 K; no feature is detected in the imaginary part. The M(H) data at 100 K and 1.8 K (but not at 300 K) show a small (1–2 × $10^{-2}$ $\mu_B$/mol) FM-like component in both orientations, which would be consistent with some canting of magnetic moments in this magnetic phase. Finally, a low-temperature upturn is observed for H || ab below ~20 K. It is not clear if this upturn could be related to the low-temperature transition predicted in our calculations, but a small Curie tail, associated with a small level of paramagnetic impurities, is a likely alternative. We must mention that the magnetic susceptibility data on $LiFe_6Ge_6$ powders [25] do not show any discernible feature below 300 K; this might be related to powder averaging or the presence of some minor secondary magnetic phase in the powder sample, as comparison of M(H) data might suggest. Anisotropic temperature-dependent resistance measurements (Fig. 4(f)) also show, for in-plane current direction, a feature at ~207 K (see derivatives dR/dT in the inset of Fig. 4(f)) is observed in magnetic measurements in the range where the possible transition between two magnetically ordered phases happens. Neither of the temperature-dependent resistivity data sets shows any feature in the 20 K temperature region, again suggesting that this is most likely associated with a small impurity-based Curie tail.

Auxiliary Mössbauer measurements (Fig. 4(g)) and the hyperfine parameters extracted from the data (even if sparse) (Fig. 4(h)) are consistent with the suggestion above that the transition below ~270 K is a transition between two magnetic phases without any significant/detected change of the hyperfine field (magnetic moment) on the $^{57}Fe$ site. These data cannot address the predicted 0.6% difference in magnetic moments on the predicted two Fe sites, since this difference is significantly less than the accuracy of the experiment. The room temperature data are consistent with those in Mantravadi et al. [25].

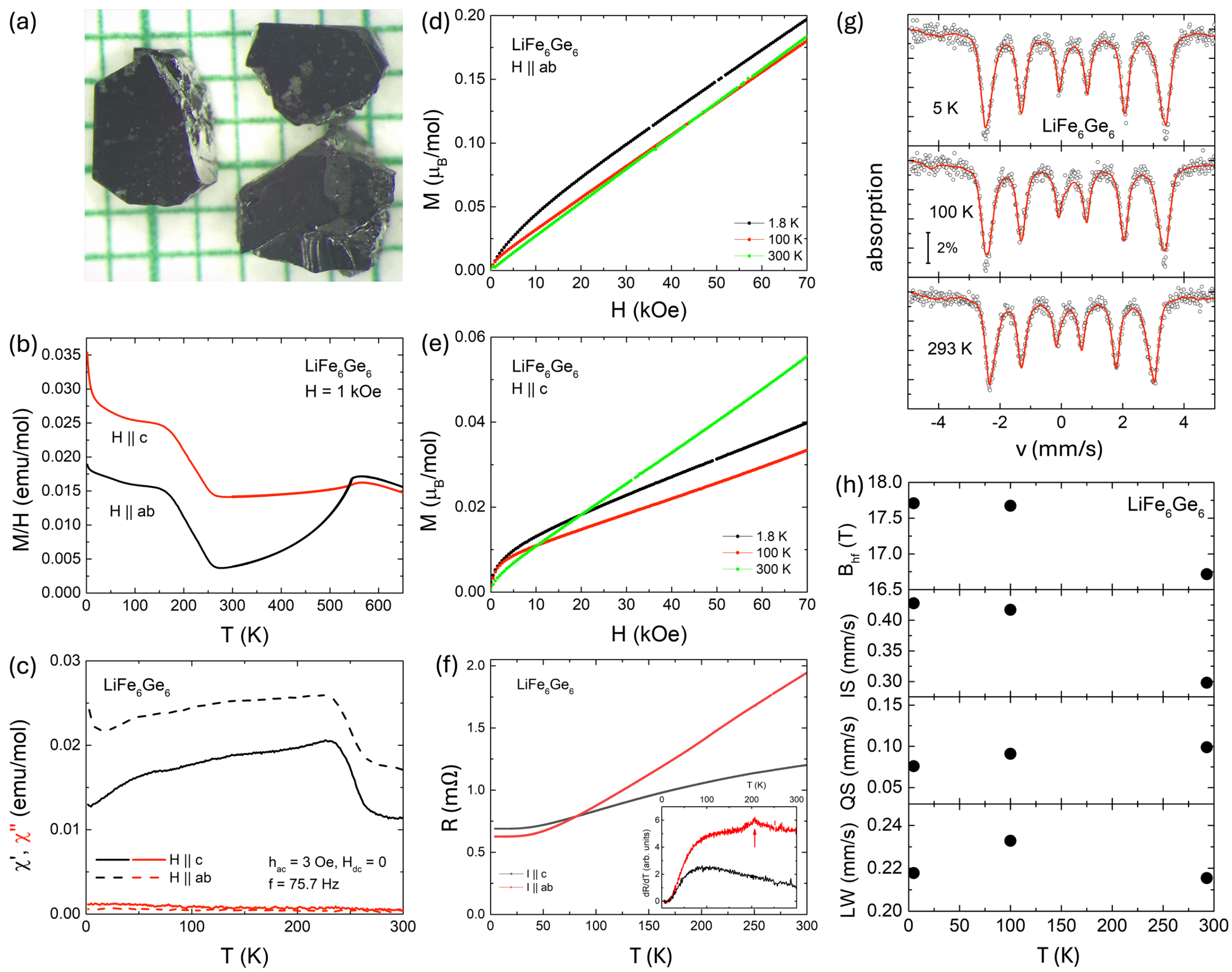

FIG. 4. (a) Photograph of $LiFe_6Ge_6$ single crystals against a mm graph paper. (b) Anisotropic dc magnetic susceptibility, M/H, of the $LiFe_6Ge_6$ single crystal measured in a 1 kOe applied field. High-temperature data (300–650 K) were shifted vertically to seamlessly join 1.8–350 K. This shift accommodates a small additional background associated with the oven heater stick. (c) Anisotropic, zero-bias dc field, ac susceptibility of $LiFe_6Ge_6$. (d)(e) Anisotropic field-dependent magnetization, M(H), measured at selected temperatures. (f) Anisotropic temperature-dependent resistance data for $LiFe_6Ge_6$. Inset: derivatives dR/dT, the arrow points to a feature possibly associated with a lower-temperature transition between two magnetic phases. (g) $^{57}$Fe Mössbauer spectra of $LiFe_6Ge_6$ measured at selected temperatures. Circles: data, lines: fits. (h) Hyperfine parameters obtained from the fits: $B_{hf}$: hyperfine field, IS: isomer shift, QS: quadrupole splitting, LW: line width.

## III. Conclusions

In summary, spin spiral calculations indicate a coplanar spin spiral ground state for $LiFe_6Ge_6$, which is slightly more stable compared to the collinear AFM state shown in Fig. 1(b) and is expected to transform to it with temperature lowering. Sizeable easy-axis magnetocrystalline anisotropy predicted for this compound suggests that the cycloidal spiral orientation is energetically preferable, with the spins lying in a plane containing the easy axis. On the other hand, spin spiral calculations for $LiFe_6Ge_4$ and $LiFe_6Ge_5$ established the stability of the collinear AFM states shown in Figs. 1(f) and 1(d), respectively. At high temperature, all three compounds are expected to exhibit collinear AFM magnetic ordering patterns, as shown in Fig. 1.

Experiments performed on $LiFe_6Ge_6$ single crystals suggest two magnetic transitions, from paramagnetic to AFM phase with the moments confined to the ab plane, at ~540 K, and then, on further cooling, a second magnetic transition below ~270 K. This transition appears to be like the one observed in $YMn_6Sn_6$ [4,40–42], which is a prototypical $AT_6X_6$ kagome material [3]. However, in our case of $LiFe_6Ge_6$, the observed Néel temperature is much higher (~345 K in $YMn_6Sn_6$) with somewhat similar, lower, and broader temperature of transition between two magnetic phases (below ~330 K in $YMn_6Sn_6$) [4,40–42]. The observed temperature sequence of magnetic phases' appearance is consistent with the calculations. Thus, our theoretical and experimental studies ultimately indicate the presence of a second magnetic transition in $LiFe_6Ge_6$ involving collinear and noncollinear magnetic phases.

The analogous $YMn_6Sn_6$ compound is known to have an intrinsic incommensurate double flat spiral state [4,7,40–42]. Magnetic fields applied to it can induce multiple transitions in the magnetization and Hall resistivity [43,44]. A highly nontrivial anomalous/topological Hall effect [4,42] was observed in its in-plane-field-induced transverse conical spiral state, which consists of an induced constant magnetic moment and a cycloidal spiral [4]. Here, the intrinsic incommensurate cycloidal spin spiral state predicted in $LiFe_6Ge_6$ inspires future exploration of such and related properties. All the materials studied in this work, $LiFe_6Ge_6$, $LiFe_6Ge_4$, and $LiFe_6Ge_5$, with similar FM kagome layers but diverse interlayer orderings, provide a promising platform for extensive unconventional or topological magnetism research by tuning various magnetic phases. Studies of related kagome materials sharing the same or similar motifs are also motivated.

Overall, the broader class of these kagome magnets, including FeGe, $AT_6X_6$, and $Fe_3Ge$ crystal, exhibit a variety of phase transitions upon cooling, e.g., collinear to noncollinear magnetic transition ($LiFe_6Ge_6$ in this study and $YMn_6Sn_6$ [4,7,40–42]), spin-reorientation transition ($TbMn_6Sn_6$ [5], $YbFe_6Ge_6$ [45,46], and $Fe_3Ge$ crystal [32]), and charge density wave (FeGe [30]).

The inter-kagome-layer spacing, the existence of the third element A in $AT_6X_6$, and how A interacts with the hosting binary compound are all likely to play critical roles in determining the magnetic behaviors in this class of kagome magnets. A future comprehensive comparison of similarity and difference among them will deepen the understanding of the interplay of their structural features, electron correlations, and various magnetic and quantum phases.

## ACKNOWLEDGMENTS

This research was supported by the U.S. Department of Energy (DOE), Office of Science, Basic Energy Sciences, Materials Science and Engineering Division. Ames National Laboratory is operated for the U.S. DOE by Iowa State University under Contract No. DE-AC02-07CH11358. The work of Z.Z. and K.B. was supported by the DOE Established Program to Stimulate Competitive Research (EPSCoR) Grant No. DE-SC0024284. Computations were performed at the High Performance Computing facility at Iowa State University and the Holland Computing Center at the University of Nebraska.

## APPENDIX

### A. Computational methods

We conducted DFT calculations using the projector augmented wave (PAW) method as implemented in the VASP package [47]. The exchange-correlation energy was treated by the Perdew-Burke-Ernzerhof (PBE) [48] generalized gradient approximation (GGA). A plane-wave basis set with a kinetic energy cutoff of 600 eV was used. A Gaussian smearing of 0.05 eV was used. The convergence thresholds were $10^{-5}$ eV for electronic self-consistency and 0.01 eV $Å^{-1}$ for ionic relaxation. A Γ-centered k-point grid of $2\pi \times 0.02$ $Å^{-1}$ spacing was used for the Brillouin zone sampling in the structural optimization, total energy, magnetic moment, and electronic density of states calculations. RKKY magnetic exchange coupling parameters were computed by using the TB2J package [49] based on localized orbitals obtained by the OpenMX package [50] with the PBE GGA functional. Spin-wave spectrum was calculated by using the SpinW code [51]. The spin spiral calculations based on the generalized Bloch theorem [39] were performed using the linear muffin-tin orbital method implementation [52], which is currently available as a part of the Questaal package [53]. The LDA exchange-correlation functional was used in the spin spiral calculations.

### B. Experimental methods

Single crystals of $LiFe_6Ge_6$ were synthesized using a self-flux solution growth method [54–56]. Small pieces of lithium (Alfa Aesar 99%), iron (Thermo Scientific 99.98%), and germanium (MSE Supplies 99.999%) were weighed with a starting composition of $Li_5Fe_{40}Ge_{55}$. The iron and germanium were weighed first, and then air-sensitive Li was added in an argon glove box. All elements were welded into a 3-cap Ta crucible [54,55]. The sealed Ta crucible was then itself sealed in a fused silica ampoule under a partial argon atmosphere. The ampoule was heated in a box furnace to 1100 °C for over 10 hours, held at the temperature for 12 hours to ensure a homogeneous melt, and then slowly cooled down to 940 °C for over 100 hours. After dwelling at 940 °C for a few hours, the excess solution was then decanted using a centrifuge [54–56]. Well-faceted hexagonal plates of single-crystalline $LiFe_6Ge_6$ were obtained, having typical dimensions of ∼4 mm × ∼4 mm with an average thickness of ∼3 mm. See Fig. 4(a) for the picture of representative crystals. The crystals are stable in air, and their stability has been confirmed by measuring powder X-ray diffraction on the same ground powder in a span of 12 hours, 24 hours, and 48 hours, with the patterns showing no change both qualitatively and quantitatively.

The phase was confirmed using a Rigaku Miniflex-II powder diffractometer using Cu Kα radiation (λ = 1.5406 Å). Single crystals were finely ground, and the powder was then mounted and measured on a single-crystal Si, zero-background sample holder, using a small amount of vacuum grease. Intensities were collected for 2θ ranging from 5° to 100° in steps of 0.01°, counting each angle for 5 seconds. The patterns were refined using GSAS-II software [57]. The peaks were matched with the peaks of the $LiFe_6Ge_6$ hexagonal structure with space group $P6/mmm$ (No. 191) [24].

Anisotropic dc and ac magnetization measurements were performed using a Quantum Design SQUID magnetometer (Magnetic Property Measurement System) MPMS3. For the 1.8–350 K temperature range, the crystal was mounted on a quartz paddle sample holder with a small amount of 7031 GE varnish. Whereas for the 300–650 K temperature range, an oven heater stick was used. In the latter case, the sample was attached to Zircar cement, and a copper foil radiation shield was used for sample homogeneity.

Temperature-dependent resistance measurements were performed in a closed-cycle cryostat (Janis SHI-950), with phosphor-bronze wires (QT-36, LakeShore, Inc.) used on the probe. Temperature was measured by a calibrated Cernox1030 sensor connected to a LakeShore 336 controller. The sample AC resistance was measured with LakeShore AC resistance bridges (models 370 and 372), with a frequency of 16.2 Hz and 3.16 mA excitation current. The electric resistance was measured for the current oriented along the c-axis and perpendicular to the c-axis, using a standard four-contact geometry. Electrical contacts with less than 2.5 Ω resistance were achieved by attaching 25 μm Pt wire to the samples with Dupont 4929N silver paint, curing at room temperature.

Auxiliary $^{57}$Fe Mössbauer spectroscopy measurements were performed using a SEE Co. conventional, constant acceleration type spectrometer in transmission geometry with a $^{57}$Co (Rh) source kept at room temperature. The absorber was prepared as powdered single crystals mixed with hBN powder for homogeneity. The driver velocity was calibrated using an α-Fe foil, and all isomer shifts are quoted relative to the α-Fe foil at room temperature. The absorber was cooled to a specific temperature using a Janis model SHI-850-5 closed-cycle refrigerator with vibration damping installed. The Mössbauer spectra were fitted using the commercial software package MossWinn [58].